\documentclass[noshowpacs,amsmath,
twocolumn,
superscriptaddress,
8pt,
aps,prb]{revtex4-2}
\pdfoutput=1
\usepackage{setspace}
\usepackage{amsmath}
\usepackage{multirow}
\usepackage{graphicx}
\usepackage[export]{adjustbox}
\usepackage{upgreek}

\usepackage{verbatim}
\usepackage{amsfonts}
\usepackage{amssymb}
\usepackage{epstopdf}
\usepackage{dcolumn}
\usepackage{xcolor}
\usepackage{verbatim}
\usepackage[version=4]{mhchem}
\usepackage{hyperref}
\usepackage{siunitx}
\usepackage{textcomp}
\setcitestyle{super}
\DeclareGraphicsExtensions{.pdf,.eps,.png,.jpg,.mps}
\begin{document}
\title{Program gain and loss for broadband soliton microcombs}

\author{Yuanlei Wang$^{1,2*}$, Xinrui Luo$^{1,*}$, Binbin Nie$^{1}$, Du Qian$^{1}$, Zhenchao Mei$^{1}$, Yanwu Liu$^{1}$, Haoyang Luo$^{1}$, Junqi Wang$^{1}$, Yiwen Yang$^{1}$, Zu-Lei Wu$^{2}$, Tianxiang Hong$^{1}$, Bei-Bei Li$^{2}$, Qihuang Gong$^{1,3,4,5}$, and Qi-Fan Yang$^{1,3,4,5\dagger}$\\
$^1$State Key Laboratory for Artificial Microstructure and Mesoscopic Physics and Frontiers Science Center for Nano-optoelectronics, School of Physics, Peking University, Beijing 100871, China\\
$^2$Beijing National Laboratory for Condensed Matter Physics, Institute of Physics, Chinese Academy of Sciences, Beijing 100190, China\\
$^3$Peking University Yangtze Delta Institute of Optoelectronics, Nantong 226010, China\\
$^4$Collaborative Innovation Center of Extreme Optics, Shanxi University, Taiyuan 030006, China\\
$^5$Hefei National Laboratory, Hefei 230088, China\\
$^{*}$These authors contributed equally to this work.\\
$^{\dagger}$Corresponding author: leonardoyoung@pku.edu.cn}

\begin{abstract}
Soliton microcombs provide compact, broadband, coherent light sources for precision metrology, spectroscopy, communications, and microwave photonics. Extending their spectral span while retaining useful output power remains challenging and often requires impractically high pump power. Existing approaches mainly tailor the dispersion and pumping conditions, but they do not exploit the coupling spectrum as a programmable aspect of soliton operation. Here we introduce a meta-coupler whose lithographically programmed coupling spectrum concentrates strong pump access near the pumped resonance while leaving most comb lines close to the intrinsic loss rate. Si$_3$N$_4$ microresonators incorporating a meta-coupler exhibit broader circulating soliton spectra, nearly twofold larger 3 dB soliton bandwidths, up to about 12 dB higher central comb-line power, and up to about fivefold greater emitted comb power, without an additional pump-power penalty. Our work unlocks gain and loss as simultaneous programmable knobs for realizing high-performance soliton microcombs.
\end{abstract}

\maketitle

\medskip

Soliton microcombs arise when a driven Kerr-nonlinear microresonator supports a stable dissipative soliton pulse\cite{leo2010temporal,herr2014temporal,Kippenberg2018}. Their combination of chip-scale integration, broadband coherence, and high repetition rate makes them attractive for precision metrology and frequency synthesis\cite{papp2014microresonator,del2016phase,spencer2018optical}, spectroscopy\cite{suh2016microresonator,dutt2018chip,suh2019searching,obrzud2019microphotonic}, communications\cite{marin2017microresonator,fulop2018high,shu2022microcomb,jorgensen2022petabit}, microwave photonics\cite{sun2024integrated,kudelin2024photonic,sun2025microcavity,jin2025microresonator,ji2025dispersive}, and ranging\cite{trocha2018ultrafast,suh2018soliton,riemensberger2020massively}. Sustaining a soliton microcomb requires a balance beyond that of traditional soliton physics: the coherent gain provided by the pump must compensate for the loss of the entire comb, and a broader comb generally requires substantially higher pump power. This creates a central dilemma for soliton microcombs: a tension between the limited power available from practical laser diodes and the broad comb span desired for many applications. Most efforts to broaden soliton microcombs have therefore focused on dispersion engineering. In particular, dispersive-wave control\cite{brasch2016photonic,pfeiffer2017octave,li2017stably,drake2019terahertz,song2024octave}, near-zero-dispersion operation\cite{anderson2022zero,xiao2023near,ji2023engineered}, and inverse-designed resonators\cite{lucas2023tailoring} have considerably expanded comb span. Other approaches modify the pumping configuration to improve efficiency\cite{yang2024efficient}, including pulsed pumping\cite{obrzud2017temporal,li2022efficiency} and coupled resonators\cite{xue2019super,helgason2023surpassing,zhu2025ultra}, which can effectively increase the available pump power. Nonetheless, these approaches increase system complexity, so a static implementation remains attractive in practice.

In contrast to the dispersion spectrum, the coupling spectrum of the bus--waveguide coupler has often been overlooked as a programmable degree of freedom. Yet it plays a central role in setting both the gain and loss conditions of soliton microcombs. In conventional couplers with little wavelength dependence, critical coupling---where the coupling rate matches the intrinsic loss---balances pump loading against intrinsic loss and is typically favorable for achieving the broadest comb. Although over-coupling can increase extracted power, the additional loss reduces the accessible comb span. This trade-off has constrained the performance of soliton microcombs in many applications.

Here we introduce a meta-coupler with a lithographically programmed coupling spectrum. For efficient soliton generation, it is designed to concentrate strong coupling near the pump while leaving most of the comb weakly coupled. Measurements show that Si$_3$N$_4$ microresonators incorporating meta-couplers support broader circulating spectra and higher emitted power.

\begin{figure*}[t!]
  \centering
  \includegraphics[max width=\linewidth]{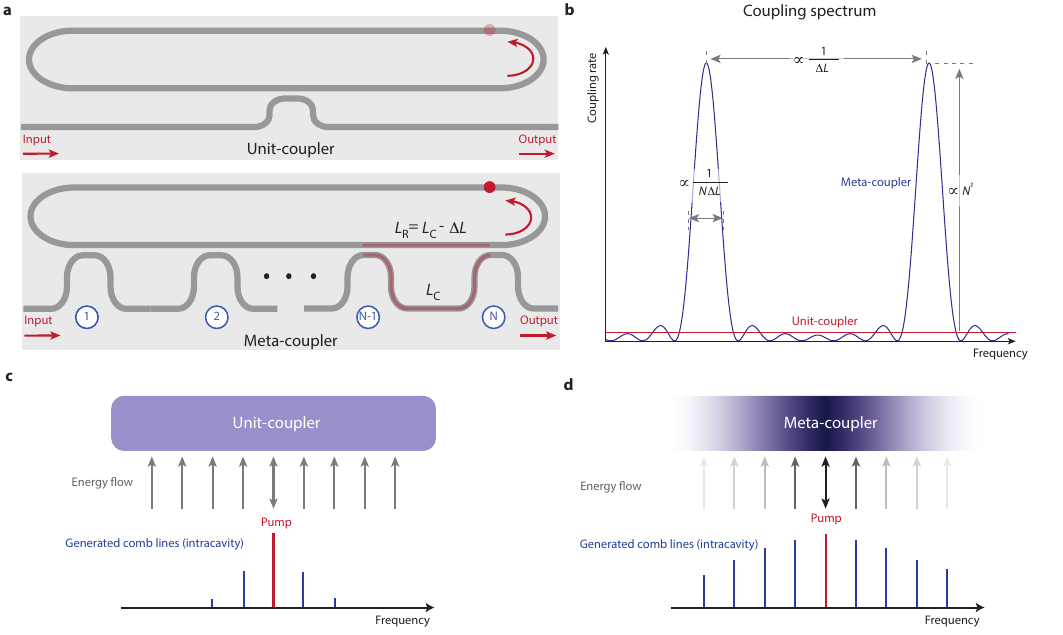}
 \caption{\textbf{Meta-coupler concept and comb dynamics.}
\textbf{a,} Schematic comparison of a unit-coupler and a meta-coupler composed of $N$ weak coupling units. The labels $1,\ldots,N$ mark the coupling units. The bus path length between adjacent units is denoted $L_C$, the corresponding resonator path length is denoted $L_R$, and their mismatch sets the interferometric phase accumulation between adjacent units.
\textbf{b,} Coupling spectrum of the meta-coupler. The typical coupling spectrum for a unit-coupler is also indicated by the red line.
\textbf{c,d,} Energy-flow picture for a unit-coupler (\textbf{c}) and a meta-coupler (\textbf{d}).}
  \label{fig:fig1}
\end{figure*}

\medskip
\noindent\textbf{Results}

\noindent\textbf{Concept.} The conventional unit-coupler and the meta-coupler are contrasted schematically in Fig.~\ref{fig:fig1}a. Whereas the unit-coupler contains only a single coupling unit, the meta-coupler contains $N$ coupling units distributed along the resonator. The bus-path length between adjacent units is denoted $L_C$, and the corresponding resonator-path length is denoted $L_R$. Their coherent superposition defines the coupling spectrum (Fig.~\ref{fig:fig1}b). Assuming that each unit is individually broadband, the interference creates a strong coupling band while suppressing coupling away from it. Increasing the number of units narrows this band and increases the coupling contrast, whereas the band spacing is set by the path-length mismatch $L_C-L_R$. The synthesized quantity is the external coupling rate $\kappa_e(\mu)$, where $\mu$ denotes the resonance number (see Methods). Once the band spacing is much larger than the measured comb span, soliton operation can be approximated as being governed by a single coupling band.

The resulting energy flow differs between the unit-coupler and the meta-coupler (Fig.~\ref{fig:fig1}c,d). With a unit-coupler, the pump and a large fraction of the comb couple to the bus waveguide with comparable strength. With a meta-coupler, strong coupling can be concentrated near the pumped resonance, while most comb lines remain closer to the intrinsic loss rate $\kappa_0$. This relaxes the power penalty for spectral broadening: strong coupling is retained where pump power is injected, but the same penalty is not imposed across most of the occupied comb. As a result, a much broader comb is expected in microresonators with meta-couplers. The corresponding coupled-mode description and heuristic scaling picture are summarized in Methods, with the full derivation provided in the Supplementary Materials.

\begin{figure*}[t!]
  \centering
  \includegraphics[max width=\linewidth]{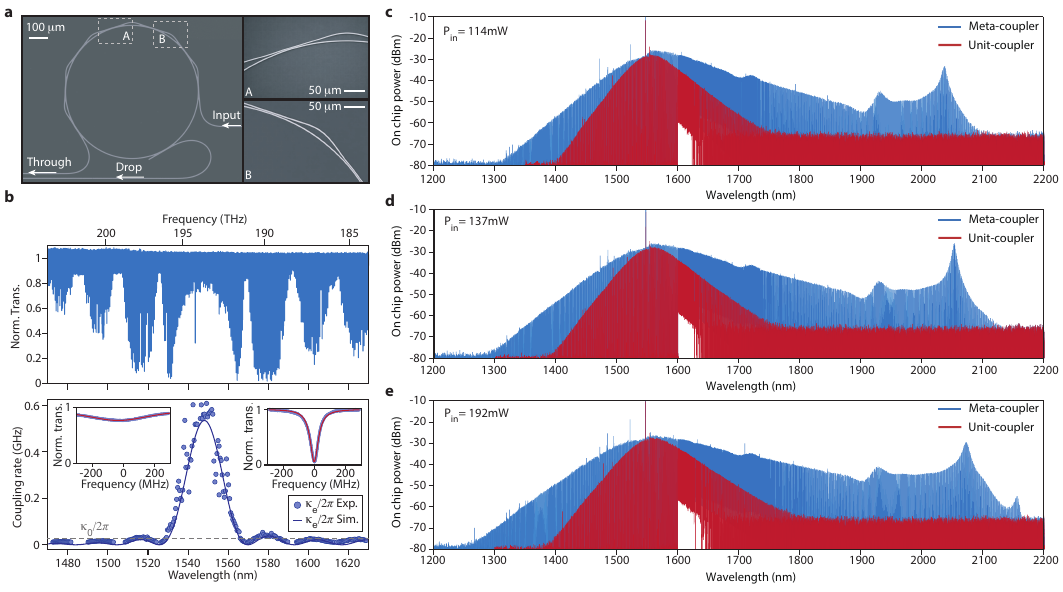}

\caption{\textbf{Device and span measurements.}
\textbf{a,} Optical micrograph of the fabricated Si$_3$N$_4$ microresonator with meta-coupler, bus waveguide and drop port. Inset A enlarges one coupling unit of the meta-coupler. Inset B shows the optical delay section between adjacent units.
\textbf{b,} Broadband transmission measurement of the resonator with a meta-coupler together with resonance fits used to extract the external coupling rate near the pump region. Insets show representative resonances inside the coupling band (left inset) and outside it (right inset). The dashed line marks the intrinsic loss rate.
\textbf{c--e,} Drop-port spectra measured from resonators with a meta-coupler (blue) and resonators with a unit-coupler (red) operated near critical coupling, at on-chip pump powers of 114, 137 and 192\,mW, respectively.}

  \label{fig:fig2}
\end{figure*}

\medskip
\noindent\textbf{Devices and span measurements.} We test the concept using Si$_3$N$_4$ microresonators fabricated on a 690-nm-thick platform (see Methods). The device includes a bus waveguide, a weakly coupled drop port, and a free spectral range of about 65~GHz (Fig.~\ref{fig:fig2}a). The bus waveguide is implemented as a meta-coupler comprising 7 units (inset A in Fig.~\ref{fig:fig2}a), with the path delay between adjacent units set to 6.87~\textmu m (inset B in Fig.~\ref{fig:fig2}a). Broadband transmission measurements, together with resonance fits, yield the external coupling rate (Fig.~\ref{fig:fig2}b), and the insets compare representative resonances inside and outside the coupling band. The extracted external coupling rate is 534~MHz near 1550~nm, whereas the intrinsic loss rate is 36.5~MHz, including the loss introduced by the drop port. The coupling at the pump wavelength is therefore about fifteen times larger than the intrinsic loss and is centred in the pump-wavelength range used for the soliton measurements below.

\begin{figure*}[t!]
  \centering
\includegraphics[max width=\linewidth]{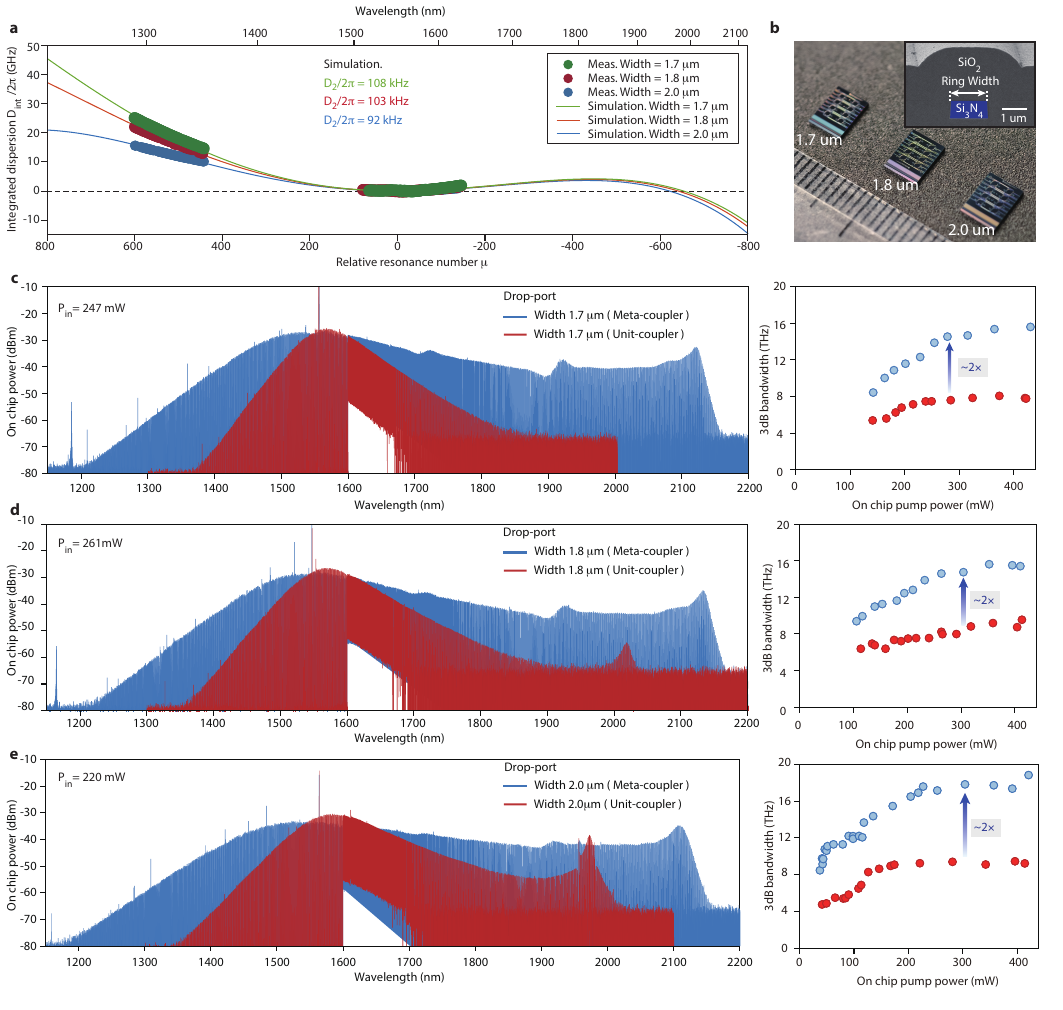}

\caption{\textbf{Universality and comb statistics.}
\textbf{a,} Measured integrated dispersion for ring widths of 1.7, 1.8 and 2.0~$\mu$m, extracted from resonances in the 1290--1350~nm and 1510--1630~nm bands, together with simulated dispersion curves.
\textbf{b,} Optical micrographs of the fabricated devices for the three ring widths. Inset, scanning-electron micrograph of the resonator cross section. All three devices share the same 690-nm Si$_3$N$_4$ film thickness; only the ring width is varied.
\textbf{c--e,} Drop-port spectra comparing resonators with a meta-coupler (blue) and resonators with a unit-coupler (red) operated near critical coupling for ring widths of 1.7, 1.8 and 2.0~$\mu$m, respectively, together with the corresponding 3 dB soliton bandwidth versus on-chip pump power.}

  \label{fig:fig3}
\end{figure*}

\begin{figure*}[t!]
  \centering
\includegraphics[max width=\linewidth]{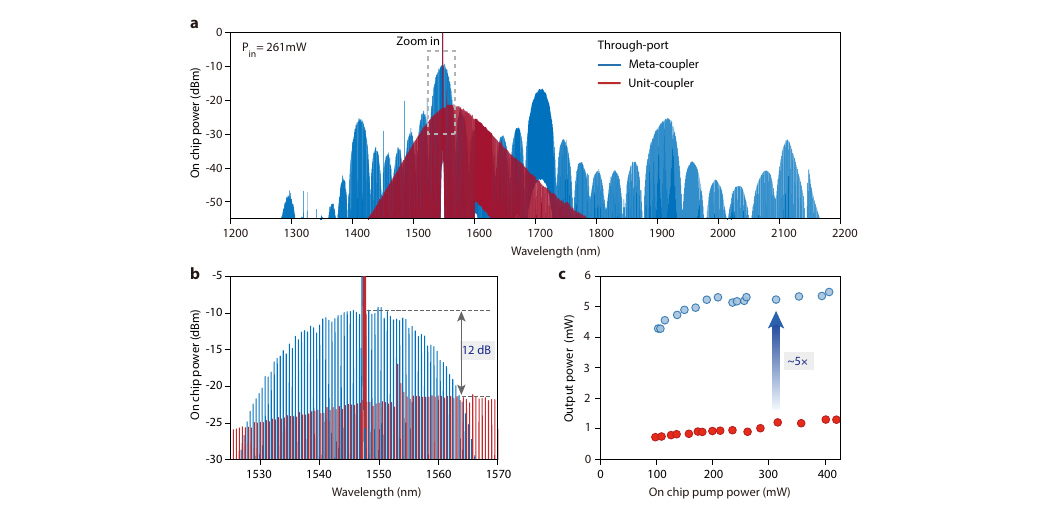}
\caption{\textbf{Output power.}
\textbf{a,} Through-port spectra of a resonator with a meta-coupler (blue) and a resonator with a unit-coupler (red) operated near critical coupling at an on-chip pump power of 261\,mW.
\textbf{b,} Enlargement of the shaded region in \textbf{a}. Within the coupling band of the meta-coupler, the power of individual comb lines is enhanced by up to $\sim$12\,dB relative to the resonator with a unit-coupler.
\textbf{c,} Emitted comb power versus on-chip pump power for $1.8\,\mu\mathrm{m}$-wide resonators with a meta-coupler and with a unit-coupler.}

  \label{fig:fig4}
\end{figure*}

For every comparison in Figs.~\ref{fig:fig2}--\ref{fig:fig4}, the resonator with a meta-coupler and the one with a unit-coupler share the same microresonator geometry, intrinsic quality factor, and dispersion; the coupler design is the only intended difference. Both devices include drop ports, which provide broadband coupling for probing the circulating soliton spectrum\cite{wang2016intracavity}. Drop-port spectra measured at on-chip pump powers of 114, 137, and 192~mW are compared in Fig.~\ref{fig:fig2}c--e for such a device pair. The pump wavelength is near 1550~nm, close to the maximum of the measured coupling band shown in Fig.~\ref{fig:fig2}b. The soliton-access protocol and on-chip pump-power calibration are described in Methods. Under these conditions, the resonator with a meta-coupler consistently supports broader soliton spectra than the resonator with a unit-coupler operated near critical coupling. The spectra also show a long-wavelength dispersive wave generated by higher-order dispersion. This feature remains prominent in the resonator with a meta-coupler because the broadened circulating spectrum extends farther into the phase-matched wing.

We next test whether the broadened span persists under different dispersion conditions. Devices with ring widths of 1.7, 1.8 and 2.0~$\mu$m are measured. The corresponding integrated dispersion, extracted from measured resonances as described in Methods, is plotted together with simulations in Fig.~\ref{fig:fig3}a. The measured and simulated integrated dispersion are in good agreement, while showing differences in second-order and higher-order dispersion across the three samples. Photographs of the fabricated chips containing the three device variants, together with a cross-sectional scanning-electron micrograph, are shown in Fig.~\ref{fig:fig3}b. These devices therefore provide distinct anomalous-dispersion conditions while preserving the same basis for comparison.

Across all three widths, the resonator with a meta-coupler produces broader drop-port spectra than the resonator with a unit-coupler at comparable on-chip pump power, and the corresponding 3 dB soliton-bandwidth traces follow the same trend over the full measured range (Fig.~\ref{fig:fig3}c--e). The increase is close to a factor of two over much of that range, which would typically require about four times higher pump power in a conventional microresonator. Dispersive waves are also observed in all three devices. The dispersive-wave peak of the resonator with a meta-coupler appears at a longer wavelength than that of the resonator with a unit-coupler because the meta-coupler devices are pumped at larger detuning, which shifts the phase-matching wavelength to the red\cite{brasch2016photonic,yang2016spatial,yi2017single}. The consistent behaviour across these devices indicates that the observed broadening is an inherent feature of the meta-coupler design.

\medskip
\noindent\textbf{Output power measurements.} Although reducing coupling-induced loss over most of the comb might appear to lower the output power, we show that meta-couplers can simultaneously provide a broader span and higher output power. The spectra acquired at the through ports are compared in Fig.~\ref{fig:fig4}a. Because the coupling band is centred near the pumped resonance, the through-port spectrum is strongly enhanced in that region. A magnified view of this band shows an increase of about 12~dB in the power of the central comb lines (Fig.~\ref{fig:fig4}b). The emitted comb power, defined here as the integrated through-port comb spectrum over the measured wavelength range with the pump line excluded, is plotted against on-chip pump power in Fig.~\ref{fig:fig4}c. Across the measured pump range, the meta-coupler yields up to about a fivefold increase relative to the resonator with a unit-coupler. We find that this behavior arises from the finite bandwidth of the meta-coupler. Although an infinitely narrow coupling band is favorable for maximizing comb span, a properly chosen coupling bandwidth can also enable efficient comb-power extraction, thereby allowing simultaneous improvements in both span and output power (see Supplementary Materials).

\medskip
\noindent\textbf{Discussion}

\noindent We have shown that programming the coupling spectrum can significantly alter the accessible properties of soliton microcombs. In the present implementation, this control is realized using a uniform array of weak coupling units. More general meta-couplers could employ nonuniform coupling strengths, spacings, or phases to synthesize flatter bands or multiple bands in analogy with fiber Bragg gratings \cite{kashyap2009fiber}, thereby enabling more flexible control over pump loading and comb output at target wavelengths. This has direct implications for several applications. Absolute frequency metrology\cite{spencer2018optical} and optical clock systems\cite{newman2019architecture} require self-referencing, so preserving comb span while enhancing extraction near the $f$--$2f$ bands is directly relevant to octave-directed comb design\cite{li2017stably,brasch2017self,drake2019terahertz}. Photonic microwave generation\cite{sun2024integrated,kudelin2024photonic,sun2025microcavity,jin2025microresonator,ji2025dispersive} and coherent communications\cite{marin2017microresonator,fulop2018high,shu2022microcomb,jorgensen2022petabit} benefit from stronger delivered power in selected comb lines, and the meta-coupler can provide both strong emission and additional spectral shaping. Dual-comb spectroscopy\cite{suh2016microresonator,dutt2018chip} can likewise benefit from efficient spectral extension and enhanced emission in target bands where chemical species exhibit strong absorption. In conclusion, the full potential of soliton microcombs can be more fully realized when both balances governing their operation—the balance between Kerr nonlinearity and dispersion, and the balance between gain and loss—are programmable.

\medskip
\noindent\textbf{Methods}

\begin{footnotesize}

\noindent\textbf{Meta-coupler design.} The present meta-coupler is formed by an array of $N$ identical weak coupling units between the bus waveguide and the resonator. If each unit provides a nominally broadband coupling rate $\kappa_e^{i}$, and if adjacent units are separated by an path mismatch $\Delta L=L_C-L_R$, where $L_C$ is the bus-path length and $L_R$ is the resonator-path length, then the external coupling spectrum is
\begin{equation}
\begin{aligned}
\kappa_e(\omega)
&=
\kappa_e^{i}
\left|
\sum_{n=0}^{N-1} e^{in\varphi}
\right|^2 \\
&=
\kappa_e^{i}
\frac{\sin^2(N\varphi/2)}{\sin^2(\varphi/2)},
\qquad
\varphi=n_{\mathrm{eff}}\omega\Delta L/c.
\end{aligned}
\label{eq:array_response}
\end{equation}
Here $\omega$ is the optical angular frequency, $\varphi$ is the interferometric phase step between adjacent units, $n_{\mathrm{eff}}$ is the effective refractive index of the guided mode, and $c$ is the speed of light. The band spacing is set by $\Delta L$, and the coupling band narrows as $N$ increases. If dispersion is neglected, substituting $\omega\approx\omega_0+\mu D_1$ into Eq.~\eqref{eq:array_response} gives the coupling spectrum as a function of resonance number $\mu$. A full derivation and the corresponding design scalings are provided in Supplementary Note~2.

\medskip
\noindent\textbf{Model.} The main-text physics is captured by the mode-dependent total loss
\begin{equation}
\kappa(\mu)=\kappa_0+\kappa_e(\mu),
\label{eq:kappa_total}
\end{equation}
where $\mu$ is the relative resonance number, $\mu=0$ denotes the pumped resonance, $\kappa_0$ is the intrinsic loss rate, and $\kappa_e(\mu)$ is the external coupling rate. In the coupled-mode description, the steady-state power balance can be written as
\begin{equation}
2\sqrt{\kappa_e(0)}\,\mathrm{Re}\!\left[s_{\mathrm{in}}a_0^*\right]
=
\sum_{\mu}\kappa(\mu)\left|a_\mu\right|^2,
\label{eq:energy_balance}
\end{equation}
where $a_\mu$ is the modal amplitude of resonance $\mu$, $a_0$ is the amplitude of the pumped resonance, and $|s_{\mathrm{in}}|^2=P_{\mathrm{in}}/\hbar\omega_p$ is the pump photon flux for pump angular frequency $\omega_p$. This relation shows that the pump term is set by the coupling rate of the pumped resonance, whereas the loss term accumulates over the occupied comb.

For a unit-coupler, the external coupling rate is approximately flat over the occupied comb, so $\kappa_e(\mu)\approx\kappa_m$ and the total loss is approximately $\kappa(\mu)\approx\kappa_0+\kappa_m$. The corresponding span scaling is therefore
\begin{equation}
B_s^{(\mathrm{uc})}\propto \frac{\sqrt{\kappa_m P_{\mathrm{in}}}}{\kappa_0+\kappa_m},
\label{eq:bandwidth_scaling_uc}
\end{equation}
which shows that the same broadband coupling that improves pump loading also increases the loss experienced by the generated comb.

To describe how a meta-coupler changes this balance, we define the fraction of intracavity comb energy inside a coupling band of width $B_c$ as
\begin{equation}
F_c \simeq \tanh\!\left[
\operatorname{arcosh}(\sqrt{2})\frac{B_c}{B_s^{(\mathrm{mc})}}
\right],
\label{eq:Fc}
\end{equation}
where $B_s^{(\mathrm{mc})}$ is the full 3 dB soliton bandwidth for a resonator with a meta-coupler. This gives the effective loss
\begin{equation}
\kappa_{\mathrm{eff}}(B_s^{(\mathrm{mc})})=\kappa_0+\kappa_mF_c,
\label{eq:keff_methods}
\end{equation}
with $\kappa_m\equiv\kappa_e(0)$, and the corresponding span scaling
\begin{equation}
B_s^{(\mathrm{mc})} \propto \sqrt{\frac{\kappa_m P_{\mathrm{in}}}
{\kappa_{\mathrm{eff}}(B_s^{(\mathrm{mc})})}} .
\label{eq:bandwidth_scaling}
\end{equation}
When the coupling band is significantly narrower than the comb, $F_c\ll 1$ and $\kappa_{\mathrm{eff}}(B_s^{(\mathrm{mc})})\approx\kappa_0$. The additional loss penalty introduced by the coupler is thereby largely removed, so increasing $\kappa_m$ can increase the span without incurring a comparable power penalty.

\medskip
\noindent\textbf{Fabrication.} The devices were fabricated on a 690-nm-thick stoichiometric Si$_3$N$_4$-on-silica platform\cite{wang2025compact}. Stress-release patterns were first defined in the SiO$_2$ undercladding and transferred by inductively coupled plasma reactive-ion etching (ICP-RIE). Stoichiometric Si$_3$N$_4$ was then deposited by low-pressure chemical vapor deposition (LPCVD) in two 350-nm steps, separated by a 180$^\circ$ wafer rotation. The waveguides and resonators were patterned by electron-beam lithography and transferred into the Si$_3$N$_4$ layer by ICP-RIE. After high-temperature annealing, the Si$_3$N$_4$ film thickness shrank to 690~nm. Finally, the silica cladding was deposited in two steps using LPCVD and plasma-enhanced chemical vapor deposition (PECVD).

\medskip
\noindent\textbf{Soliton generation.} A continuous-wave laser near 1550~nm was amplified, polarization-adjusted, and launched into the bus waveguide to excite the target resonance family. Stable soliton states were accessed by rapidly scanning the pump frequency across the resonance with a single-sideband modulator\cite{stone2018thermal}. Optical spectra were recorded from both the through port and the drop port using an optical spectrum analyzer. The on-chip pump power was obtained from calibrated insertion-loss measurements referenced to the chip facets.

\medskip
\noindent\textbf{Dispersion measurement.} The resonant frequencies of the target mode family were measured over the characterized wavelength range using a tunable continuous-wave laser calibrated by a Mach--Zehnder interferometer. The integrated dispersion was then obtained according to
\begin{equation}
D_{\mathrm{int}}(\mu)=\omega_\mu-(\omega_0+D_1\mu),
\label{eq:Dint}
\end{equation}
where $\omega_\mu$ is the angular frequency of resonance $\mu$, $\omega_0$ is the angular frequency of the pumped resonance, and $D_1/2\pi$ is the free spectral range. Polynomial fitting of the measured $D_{\mathrm{int}}$ yields the second-order dispersion used in the main text, in good agreement with simulation.

\end{footnotesize}
\medskip
\noindent\textbf{Acknowledgments}

\begin{footnotesize}
\noindent This work was supported by the Quantum Science and Technology-National Science and Technology Major Project (2021ZD0301500). The authors thank Jincheng Li, Xin Zhou, Jian-Fei Liu, Xuening Cao, and Zhi-Gang Hu for assistance in fabrication. The fabrication in this work was supported by the Peking University Nano-Optoelectronic Fabrication Center, Micro/nano Fabrication Laboratory of Synergetic Extreme Condition User Facility (SECUF), Songshan Lake Materials Laboratory, and the Advanced Photonics Integrated Center of Peking University. 
\end{footnotesize}
\medskip

\noindent\textbf{Author contributions} 

\begin{footnotesize}
\noindent Experiments were conceived and designed by Y.W., X.L., and Q.-F.Y. Device design and fabrication were carried out by Y.W. with assistance from H.L., J.W., X.L., Z.M., Z.-L.W., and T.H. Measurements and data analysis were performed by Y.W., X.L., and Q.-F.Y. with assistance from D.Q., Z.M., Y.L., and Y.Y. Numerical simulations and analytical modelling were performed by X.L., Y.W., and Q.-F.Y., with assistance from B.N. B.-B.L. contributed to discussions. Q.-F.Y. supervised the project.
All authors participated in preparing the manuscript.
\end{footnotesize}

\medskip

\noindent\textbf{Competing interests}

\begin{footnotesize}
\noindent The authors declare no competing interests.
\end{footnotesize}

\bibliography{refbase.bib}

\medskip

\noindent\textbf{Data availability}

\begin{footnotesize}
\noindent The data that support the plot within this paper and other findings of this study are available upon publication. 
\end{footnotesize}

\medskip

\noindent\textbf{Code availability}

\begin{footnotesize}
\noindent The codes that support the findings of this study are available upon publication. 
\end{footnotesize}

\medskip

\noindent\textbf{Additional information}

\begin{footnotesize}
\noindent Correspondence and requests for materials should be addressed to Q-F.Y.
\end{footnotesize}

\end{document}


\title{Program gain and loss for broadband soliton microcombs}

\author{Yuanlei Wang$^{1,2*}$, Xinrui Luo$^{1,*}$, Binbin Nie$^{1}$, Du Qian$^{1}$, Zhenchao Mei$^{1}$, Yanwu Liu$^{1}$, Haoyang Luo$^{1}$, Junqi Wang$^{1}$, Yiwen Yang$^{1}$, Zu-Lei Wu$^{2}$, Tianxiang Hong$^{1}$, Bei-Bei Li$^{2}$, Qihuang Gong$^{1,3,4,5}$, and Qi-Fan Yang$^{1,3,4,5\dagger}$\\
$^1$State Key Laboratory for Artificial Microstructure and Mesoscopic Physics and Frontiers Science Center for Nano-optoelectronics, School of Physics, Peking University, Beijing 100871, China\\
$^2$Beijing National Laboratory for Condensed Matter Physics, Institute of Physics, Chinese Academy of Sciences, Beijing 100190, China\\
$^3$Peking University Yangtze Delta Institute of Optoelectronics, Nantong 226010, China\\
$^4$Collaborative Innovation Center of Extreme Optics, Shanxi University, Taiyuan 030006, China\\
$^5$Hefei National Laboratory, Hefei 230088, China\\
$^{*}$These authors contributed equally to this work.\\
$^{\dagger}$Corresponding author: leonardoyoung@pku.edu.cn}

\maketitle
\tableofcontents
\newpage

\section{Theory}

\subsection{Generalized LLE and gain and loss balance}

To describe the mode-dependent loss discussed in the main text, we start from a generalized Lugiato--Lefever equation\cite{lugiato1987spatial,chembo2013spatiotemporal,coen2013modeling,godey2014stability} and derive the corresponding gain and loss balance by expanding the intracavity field as
\begin{equation}
A(\phi,T)=\sum_{\mu=-\infty}^{\infty} a_\mu(T)e^{i\mu\phi}
\qquad
a_\mu(T)=\frac{1}{2\pi}\int_{0}^{2\pi}A(\phi,T)e^{-i\mu\phi}\mathrm{d}\phi
\label{eq:S_Fourier_def}
\end{equation}
Here $T$ is the slow time, $\phi$ is the angular coordinate in the moving frame, and $\mu$ is the resonance number relative to the pumped resonance. The pumped resonance corresponds to $\mu=0$. The slowly varying field amplitude is $A(\phi,T)$, and $a_\mu(T)$ is the corresponding modal amplitude. With this convention, Parseval's relation gives,
\begin{equation}
\frac{1}{2\pi}\int_{0}^{2\pi}|A|^2\mathrm{d}\phi=\sum_\mu |a_\mu|^2.
\label{eq:S_Parseval}
\end{equation}
so that $|a_\mu|^2$ gives the intracavity energy in the resonance indexed by $\mu$, and the right-hand side gives the total intracavity energy.

To include mode-dependent loss, we define the operator,
\begin{equation}
\hat{\kappa}A\equiv \sum_\mu \kappa(\mu)a_\mu e^{i\mu\phi}
\qquad
\kappa(\mu)=\kappa_{\mathrm e}(\mu)+\kappa_0(\mu)
\label{eq:S_kappa_operator}
\end{equation}
where $\kappa_0(\mu)$ is the intrinsic loss rate of resonance $\mu$, and $\kappa_{\mathrm e}(\mu)$ is the external coupling rate to the bus waveguide. Throughout this section, all detunings, decay rates, coupling rates, and dispersion coefficients are expressed in angular-frequency units.
The generalized Lugiato--Lefever equation is then written as,
\begin{equation}
\frac{\partial A}{\partial T}
=-\left(\frac{\hat{\kappa}}{2}+i\delta\omega\right)A
+i\frac{D_2}{2}\frac{\partial^2 A}{\partial \phi^2}
+ig|A|^2A
+\sqrt{\kappa_{\mathrm e}(0)P_{\mathrm{in}}}
\label{eq:S_LLE}
\end{equation}
With Kerr coefficient
\begin{equation}
g=\frac{\omega_0 c n_2}{n_0^2 V_{\mathrm{eff}}}
\label{eq:S_g_def}
\end{equation}
where $\delta\omega=\omega_0-\omega_p$ is the pump-resonance detuning, with
$\omega_0$ and $\omega_p$ denoting the angular frequencies of the pumped
cavity resonance and the pump laser, respectively. $D_2$ is the second-order
dispersion coefficient of the resonator. In Eq.~\eqref{eq:S_g_def}, $c$ is
the vacuum speed of light, $n_2$ is the Kerr nonlinear index, $n_0$ is the
refractive index at $\omega_0$, and $V_{\mathrm{eff}}$ is the effective
optical mode volume. $P_{\mathrm{in}}$ is the on-chip pump power. The last
term in Eq.~\eqref{eq:S_LLE} is independent of $\phi$ and therefore drives
only the pumped resonance, $\mu=0$, with coupling rate $\kappa_{\mathrm e}(0)$.

We next define the normalized intracavity energy
\begin{equation}
U(T)\equiv \frac{1}{2\pi}\int_{0}^{2\pi}|A|^2\mathrm{d}\phi=\sum_\mu |a_\mu|^2.
\label{eq:S_U_def}
\end{equation}
Differentiating $U$ and using Eq.~(\ref{eq:S_LLE}), the detuning, dispersion, and Kerr terms are purely reactive and do not contribute to $\mathrm{d}U/\mathrm{d}T$ under periodic boundary conditions. One obtains
\begin{equation}
\frac{\mathrm{d}U}{\mathrm{d}T}
=-\sum_\mu \kappa(\mu)|a_\mu|^2
+2\sqrt{\kappa_{\mathrm e}(0)P_{\mathrm{in}}}\mathrm{Re}(a_0).
\label{eq:S_U_rate}
\end{equation}
In steady state $\mathrm{d}U/\mathrm{d}T=0$, yielding the gain and loss balance used in the main text
\begin{equation}
2\sqrt{\kappa_{\mathrm e}(0)P_{\mathrm{in}}}\mathrm{Re}(a_0)
=\sum_\mu \kappa(\mu)|a_\mu|^2.
\label{eq:S_balance}
\end{equation}
The pumping term is maximized when $\arg(a_0)=0$ so that $\mathrm{Re}(a_0)\le |a_0|$.

To obtain closed-form scalings, we work in the high-finesse large-red-detuning regime $\delta\omega\gg \kappa(\mu)$, where pump and loss act as perturbations and the soliton envelope is well approximated by the conservative stationary NLSE\cite{leo2010temporal,herr2014Temporal,yi2015soliton}. Neglecting the continuous-wave background and retaining only second-order dispersion and Kerr nonlinearity gives
\begin{equation}
A(\phi)\simeq A_{\mathrm s}\mathrm{sech}\!\left(\frac{\phi-\phi_0}{\tau}\right)e^{i\psi}
\qquad
A_{\mathrm s}=\sqrt{\frac{2\delta\omega}{g}}
\qquad
\tau=\sqrt{\frac{D_2}{2\delta\omega}}
\label{eq:S_soliton_ansatz}
\end{equation}
valid for anomalous dispersion $D_2>0$, where $\phi_0$ is the soliton center and $\psi$ is the overall soliton phase. For broadband solitons $\tau\ll 2\pi$, extending the Fourier integral to $\pm\infty$ yields
\begin{equation}
a_\mu \simeq \frac{A_{\mathrm s}\tau}{2}\mathrm{sech}\!\left(\frac{\pi\mu\tau}{2}\right)e^{i\psi}e^{-i\mu\phi_0}
\qquad
|a_0|\simeq \frac{A_{\mathrm s}\tau}{2}=\frac{1}{2}\sqrt{\frac{D_2}{g}}.
\label{eq:S_amu_sech}
\end{equation}
The total intracavity energy satisfies
\begin{equation}
\sum_\mu |a_\mu|^2
\simeq \int_{-\infty}^{\infty}|a_0|^2\mathrm{sech}^2\!\left(\frac{\pi\mu\tau}{2}\right)\mathrm{d}\mu
=\frac{1}{\pi g}\sqrt{2D_2\delta\omega}.
\label{eq:S_sum_energy}
\end{equation}

To connect this envelope to an experimentally accessible spectral scale, we define the full 3 dB soliton bandwidth in angular-frequency units as $B_s$. The 3 dB point satisfies $\mathrm{sech}^2(\pi\mu\tau/2)=1/2$, which gives
\begin{equation}
\mu_{3\mathrm{dB}}=\frac{2}{\pi\tau}\mathrm{arcosh}(\sqrt{2}).
\label{eq:S_mu3dB}
\end{equation}
With angular free spectral range $D_1$, the full 3 dB span is
\begin{equation}
B_s=2\mu_{3\mathrm{dB}}D_1
=\frac{4D_1}{\pi}\mathrm{arcosh}(\sqrt{2})\sqrt{\frac{2\delta\omega}{D_2}}.
\label{eq:S_Bs_detuning}
\end{equation}
When bandwidths are quoted numerically below, we normalize the angular-frequency bandwidths by $D_1$.

We then define the peak coupling rate at the pumped resonance as $\kappa_m\equiv \kappa_{\mathrm e}(0)$. Using Eq.~(\ref{eq:S_amu_sech}), the maximized pumping term in Eq.~(\ref{eq:S_balance}) scales as
\begin{equation}
\left[2\sqrt{\kappa_m P_{\mathrm{in}}}\mathrm{Re}(a_0)\right]_{\max}
=2\sqrt{\kappa_m P_{\mathrm{in}}}|a_0|
\simeq \sqrt{\frac{D_2}{g}}\sqrt{\kappa_m P_{\mathrm{in}}}.
\label{eq:S_pump_max}
\end{equation}

\subsection{Span scaling laws}

We first consider the conventional unit-coupler limit, for which $\kappa_{\mathrm e}(\mu)\approx \kappa_m$ over the soliton bandwidth and $\kappa_0(\mu)\approx \kappa_0$, so that $\kappa(\mu)\approx \kappa_0+\kappa_m$. Substituting Eqs.~(\ref{eq:S_pump_max}) and (\ref{eq:S_sum_energy}) into Eq.~(\ref{eq:S_balance}) and eliminating $\delta\omega$ using Eq.~(\ref{eq:S_Bs_detuning}) gives
\begin{equation}
B_s^{(\mathrm{uc})}
=4D_1\mathrm{arcosh}(\sqrt{2})\sqrt{\frac{g}{D_2}}
\frac{\sqrt{\kappa_m P_{\mathrm{in}}}}{\kappa_0+\kappa_m}.
\label{eq:S_Bs_bb}
\end{equation}
This expression yields the unit-coupler optimum at critical coupling $\kappa_m=\kappa_0$ for fixed $P_{\mathrm{in}}$. The resulting optimal 3 dB soliton bandwidth is
\begin{equation}
B_{s,\mathrm{opt}}^{(\mathrm{uc})}
=2D_1\mathrm{arcosh}(\sqrt{2})\sqrt{\frac{gP_{\mathrm{in}}}{D_2\kappa_0}}.
\label{eq:S_Bs_uc_opt}
\end{equation}

We next introduce a reduced square-band model as an analytical approximation to the meta-coupler used in the main text. This model retains the peak coupling rate at the pumped resonance and the width of the coupling band while suppressing weaker spectral structure that is not needed for the gain and loss balance developed here. We therefore model the coupling spectrum by a coupling band of angular-frequency width $B_c$ and peak coupling $\kappa_m$
\begin{equation}
\kappa_{\mathrm e}(\mu)=
\begin{cases}
\kappa_m & |\mu D_1|\le B_c/2 \\
0 & |\mu D_1|>B_c/2
\end{cases}
\label{eq:S_square_coupler}
\end{equation}
with uniform intrinsic loss $\kappa_0$. Using the broadband-soliton approximation of Eq.~(\ref{eq:S_amu_sech}) and replacing the modal sum by a continuous integral, the fraction of intracavity energy within the coupling band is
\begin{equation}
F_c \equiv \frac{\sum_{|\mu D_1|\le B_c/2}|a_\mu|^2}{\sum_\mu |a_\mu|^2}
\simeq \tanh\!\left(\mathrm{arcosh}(\sqrt{2})\frac{B_c}{B_s}\right)
\label{eq:S_Fc}
\end{equation}
so that the collective loss can be written as an effective loss rate
\begin{equation}
\kappa_{\mathrm{eff}}(B_s)=\kappa_0+\kappa_m F_c.
\label{eq:S_kappa_eff}
\end{equation}
Replacing $\kappa_0+\kappa_m$ by $\kappa_{\mathrm{eff}}(B_s)$ in Eq.~(\ref{eq:S_Bs_bb}) gives the corresponding implicit expression
\begin{equation}
B_s^{(\mathrm{mc})}
=4D_1\mathrm{arcosh}(\sqrt{2})\sqrt{\frac{g}{D_2}}
\frac{\sqrt{\kappa_m P_{\mathrm{in}}}}
{\kappa_0+\kappa_m\tanh\!\left(\mathrm{arcosh}(\sqrt{2})\frac{B_c}{B_s^{(\mathrm{mc})}}\right)}.
\label{eq:S_Bs_mc_exact}
\end{equation}
which is the square-band-coupler bandwidth relation used below. It is exact within the reduced square-band model once the soliton envelope is approximated by Eq.~(\ref{eq:S_amu_sech}).

Comparison with the unit coupler is most transparent at a fixed target bandwidth. Rearranging Eq.~(\ref{eq:S_Bs_mc_exact}) gives the pump power required to sustain a square-band-coupler soliton bandwidth $B_s$,
\begin{equation}
P_{\mathrm{in}}^{(\mathrm{mc})}(B_s)
=\frac{D_2}{16D_1^2\mathrm{arcosh}^2(\sqrt{2})\,g\,\kappa_m}
B_s^2
\left[\kappa_0+\kappa_m\tanh\!\left(\mathrm{arcosh}(\sqrt{2})\frac{B_c}{B_s}\right)\right]^2.
\label{eq:S_Pin_mc}
\end{equation}
For the unit coupler at critical coupling, Eq.~(\ref{eq:S_Bs_uc_opt}) gives
\begin{equation}
P_{\mathrm{in,crit}}^{(\mathrm{uc})}(B_s)
=\frac{D_2\kappa_0}{4D_1^2\mathrm{arcosh}^2(\sqrt{2})\,g}
B_s^2.
\label{eq:S_Pin_uc_crit}
\end{equation}
The corresponding pump-power ratio is therefore
\begin{equation}
\frac{P_{\mathrm{in}}^{(\mathrm{mc})}(B_s)}{P_{\mathrm{in,crit}}^{(\mathrm{uc})}(B_s)}
=\frac{\left[\kappa_0+\kappa_m\tanh\!\left(\mathrm{arcosh}(\sqrt{2})\frac{B_c}{B_s}\right)\right]^2}
{4\kappa_0\kappa_m}.
\label{eq:S_P_ratio}
\end{equation}
This expression makes the physical difference between the two couplers explicit. In the unit coupler, the full soliton spectrum incurs the external-coupling contribution $\kappa_m$, so the bandwidth is limited by the total loss rate $\kappa_0+\kappa_m$. In the square-band coupler, that contribution is weighted only by the spectral fraction inside the coupling band, so strong pump coupling can coexist with reduced collective loss across most comb lines.

The limiting cases follow directly from Eqs.~(\ref{eq:S_Bs_mc_exact}) and (\ref{eq:S_P_ratio}). If $B_c\gg B_s$, then $F_c\to 1$ and $\kappa_{\mathrm{eff}}(B_s)\to \kappa_0+\kappa_m$ so Eq.~(\ref{eq:S_Bs_mc_exact}) reduces to the unit-coupler case. If $B_c\ll B_s$, then using $\tanh z\simeq z$ for $z\ll 1$ gives
\begin{equation}
F_c\simeq \mathrm{arcosh}(\sqrt{2})\frac{B_c}{B_s}
\qquad
\kappa_{\mathrm{eff}}(B_s)\simeq \kappa_0+\kappa_m\mathrm{arcosh}(\sqrt{2})\frac{B_c}{B_s}.
\label{eq:S_smallBc}
\end{equation}
In the limit where sidelobe contributions to the external coupling rate are negligible and the coupling band is vanishingly narrow, $\kappa_{\mathrm{eff}}(B_s)\rightarrow \kappa_0$ and Eq.~(\ref{eq:S_Bs_mc_exact}) approaches $B_s\propto \sqrt{\kappa_m P_{\mathrm{in}}}/\kappa_0$, while Eq.~(\ref{eq:S_P_ratio}) approaches $\kappa_0/(4\kappa_m)$. The square-band coupler therefore preserves the same square-root dependence on $\kappa_m P_{\mathrm{in}}$ as the unit coupler, but with a smaller effective loss because only the resonances inside the coupling band incur strong external coupling.

\subsection{Power scaling laws}

To connect the same reduced model to the emitted comb power discussed in the main text, we combine Eqs.~(\ref{eq:S_sum_energy}) and (\ref{eq:S_Bs_detuning}) to write the total intracavity energy as
\begin{equation}
\sum_\mu |a_\mu|^2
\simeq
\frac{D_2}{4D_1\mathrm{arcosh}(\sqrt{2})\,g}B_s.
\label{eq:S_U_Bs}
\end{equation}
For comb lines with $\mu\neq 0$, the standard input--output relation at the through port gives the output-field amplitude $s_{\mathrm{out},\mu}$ in resonance $\mu$,
\begin{equation}
s_{\mathrm{out},\mu}=-\sqrt{\kappa_{\mathrm e}(\mu)}\,a_\mu,
\label{eq:S_sout}
\end{equation}
so the through-port emitted comb power, defined in the same way as in the main text by excluding the pump line, is
\begin{equation}
P_{\mathrm{emit}}
\equiv
\sum_{\mu\neq 0}\kappa_{\mathrm e}(\mu)|a_\mu|^2.
\label{eq:S_Pemit_def}
\end{equation}
In the broadband-soliton regime, the pumped resonance contributes only a subleading fraction of the total intracavity energy and is excluded from Eq.~(\ref{eq:S_Pemit_def}), so for the square-band coupler this becomes
\begin{equation}
P_{\mathrm{emit}}^{(\mathrm{mc})}
\simeq
\kappa_m F_c\sum_\mu |a_\mu|^2
\simeq
\frac{D_2\kappa_m}{4D_1\mathrm{arcosh}(\sqrt{2})\,g}
B_s
\tanh\!\left(\mathrm{arcosh}(\sqrt{2})\frac{B_c}{B_s}\right).
\label{eq:S_Pemit_mc_Bs}
\end{equation}
Substituting Eq.~(\ref{eq:S_Bs_mc_exact}) into Eq.~(\ref{eq:S_Pemit_mc_Bs}) gives the corresponding emitted-power relation within the same reduced model,
\begin{equation}
P_{\mathrm{emit}}^{(\mathrm{mc})}
=
\sqrt{\frac{D_2}{g}}
\frac{
\kappa_m
\tanh\!\left(\mathrm{arcosh}(\sqrt{2})\frac{B_c}{B_s^{(\mathrm{mc})}}\right)
\sqrt{\kappa_m P_{\mathrm{in}}}
}{
\kappa_0+\kappa_m
\tanh\!\left(\mathrm{arcosh}(\sqrt{2})\frac{B_c}{B_s^{(\mathrm{mc})}}\right)
}.
\label{eq:S_Pemit_mc_exact}
\end{equation}
For the unit coupler at critical coupling, $\kappa_m=\kappa_0$, Eq.~(\ref{eq:S_Bs_uc_opt}) gives the corresponding reference emitted comb power
\begin{equation}
P_{\mathrm{emit,crit}}^{(\mathrm{uc})}
=
\frac{1}{2}\sqrt{\frac{D_2\kappa_0P_{\mathrm{in}}}{g}}.
\label{eq:S_Pemit_uc_crit}
\end{equation}

The simultaneous scaling of span and emitted comb power becomes most transparent by defining
\begin{equation}
x\equiv \frac{\kappa_m}{\kappa_0},
\qquad
y\equiv F_c=
\tanh\!\left(\mathrm{arcosh}(\sqrt{2})\frac{B_c}{B_s^{(\mathrm{mc})}}\right).
\label{eq:S_xy_def}
\end{equation}
Equations~(\ref{eq:S_Bs_mc_exact}), (\ref{eq:S_Bs_uc_opt}), and (\ref{eq:S_Pemit_mc_exact}) then give
\begin{equation}
\frac{B_s^{(\mathrm{mc})}}{B_{s,\mathrm{opt}}^{(\mathrm{uc})}}
=
\frac{2\sqrt{x}}{1+xy},
\label{eq:S_B_ratio_xy}
\end{equation}
\begin{equation}
\frac{P_{\mathrm{emit}}^{(\mathrm{mc})}}{P_{\mathrm{emit,crit}}^{(\mathrm{uc})}}
=
\frac{2x^{3/2}y}{1+xy}
=
xy\,
\frac{B_s^{(\mathrm{mc})}}{B_{s,\mathrm{opt}}^{(\mathrm{uc})}}.
\label{eq:S_Pemit_ratio_xy}
\end{equation}
Here $x$ measures the enhancement of the peak pump-resonance coupling relative to the intrinsic loss, and $0<y<1$ is the spectral-overlap factor between the soliton and the coupling band.
For $x>1$, both ratios exceed unity when
\begin{equation}
\frac{1}{x(2\sqrt{x}-1)}
<
y
<
\frac{2\sqrt{x}-1}{x}.
\label{eq:S_y_window}
\end{equation}
Equation~(\ref{eq:S_y_window}) identifies the window of the effective overlap factor $y=F_c$ for which the meta-coupler, approximated here by the square-band coupler, provides both larger soliton bandwidth and larger emitted comb power than a unit coupler near critical coupling at the same pump power. In the square-band model, this corresponds to a finite range of coupling-band widths $B_c$. In the opposite limits, the tradeoff is recovered: when the coupling band is too narrow, $y\rightarrow 0$ and emitted comb power vanishes even though the span enhancement is strongest, whereas when the coupling band is too broad, $y\rightarrow 1$ and the span enhancement disappears.

Choosing $xy\simeq 1$ gives the balanced condition
\begin{equation}
\frac{B_s^{(\mathrm{mc})}}{B_{s,\mathrm{opt}}^{(\mathrm{uc})}}=
\frac{P_{\mathrm{emit}}^{(\mathrm{mc})}}{P_{\mathrm{emit,crit}}^{(\mathrm{uc})}}
\simeq
\sqrt{x},
\qquad
y\simeq \frac{1}{x},
\qquad
\frac{B_c}{B_s^{(\mathrm{mc})}}
\simeq
\frac{\operatorname{artanh}(1/x)}{\operatorname{arcosh}(\sqrt{2})}.
\label{eq:S_balanced_xy}
\end{equation}
At the larger coupling rates accessed experimentally, substantially larger simultaneous enhancement follows. For $x=12$, $16$, and $20$, the balanced condition gives $y\simeq 0.083$, $0.063$, and $0.050$, respectively, with $B_c/B_s^{(\mathrm{mc})}\simeq 0.095$, $0.071$, and $0.057$. For a representative broadband state with $B_s^{(\mathrm{mc})}/D_1\approx 200$, these values correspond to $B_c/D_1\approx 19$, $14$, and $11$, together with $\kappa_m\approx 12\kappa_0$, $16\kappa_0$, and $20\kappa_0$. The resulting simultaneous enhancement factors are
\begin{equation}
\frac{B_s^{(\mathrm{mc})}}{B_{s,\mathrm{opt}}^{(\mathrm{uc})}}=
\frac{P_{\mathrm{emit}}^{(\mathrm{mc})}}{P_{\mathrm{emit,crit}}^{(\mathrm{uc})}}
\approx
3.46,\;4.00,\;4.47,
\end{equation}
respectively. A representative range for approximately fourfold enhancement in both span and emitted comb power is therefore $x\approx 16$--$20$ with $y\approx 0.05$--$0.06$, corresponding to $\kappa_m\approx (16$--$20)\kappa_0$ and $B_c/D_1\approx 11$--$15$ for $B_s^{(\mathrm{mc})}/D_1\approx 200$. If emitted comb power is prioritized over the balanced condition, one may choose $y$ slightly above $1/x$ while remaining inside the window of Eq.~(\ref{eq:S_y_window}); for example, $x=20$ and $y=0.08$ give $B_s^{(\mathrm{mc})}/B_{s,\mathrm{opt}}^{(\mathrm{uc})}\approx 3.44$ and $P_{\mathrm{emit}}^{(\mathrm{mc})}/P_{\mathrm{emit,crit}}^{(\mathrm{uc})}\approx 5.50$, with $B_c/B_s^{(\mathrm{mc})}\approx 0.091$, i.e. $B_c/D_1\approx 18$ for $B_s^{(\mathrm{mc})}/D_1\approx 200$. For a slightly wider coupling band, $x=20$ and $y=0.10$ give $B_s^{(\mathrm{mc})}/B_{s,\mathrm{opt}}^{(\mathrm{uc})}\approx 2.98$ and $P_{\mathrm{emit}}^{(\mathrm{mc})}/P_{\mathrm{emit,crit}}^{(\mathrm{uc})}\approx 5.96$, with $B_c/D_1\approx 23$.

\section{Numerical simulations}

\subsection{Numerical simulation of broadened comb span}

Figure~S1 compares the scaling derived above with numerical simulations of unit-coupler and square-band coupler. In Fig.~S1a, the unit-coupler reference is modeled by taking $B_c\rightarrow\infty$ with $\kappa_m=\kappa_0$, whereas the square-band coupler uses $B_c=10D_1$ and peak coupling rates $\kappa_m=2.5\kappa_0$ at the pumped resonance, with the same intrinsic loss rate $\kappa_0$. Figure~S1b compares the intracavity-power evolution and the corresponding intracavity field evolution during the detuning scan at fixed on-chip pump power, showing that the square-band coupler broadens the detuning range over which solitons remain accessible. Figure~S1c shows the theoretical phase diagrams in pump power and detuning. Figure~S1d compares the resulting 3 dB soliton bandwidth versus on-chip pump power. Under the conditions marked in Fig.~S1c, the square-band coupler reaches a comparable 3 dB soliton bandwidth with approximately fivefold lower pump power than the resonator with a unit coupler, consistent with the reduced collective-loss scaling.

\begin{figure*}[htbp]
  \centering
  \includegraphics{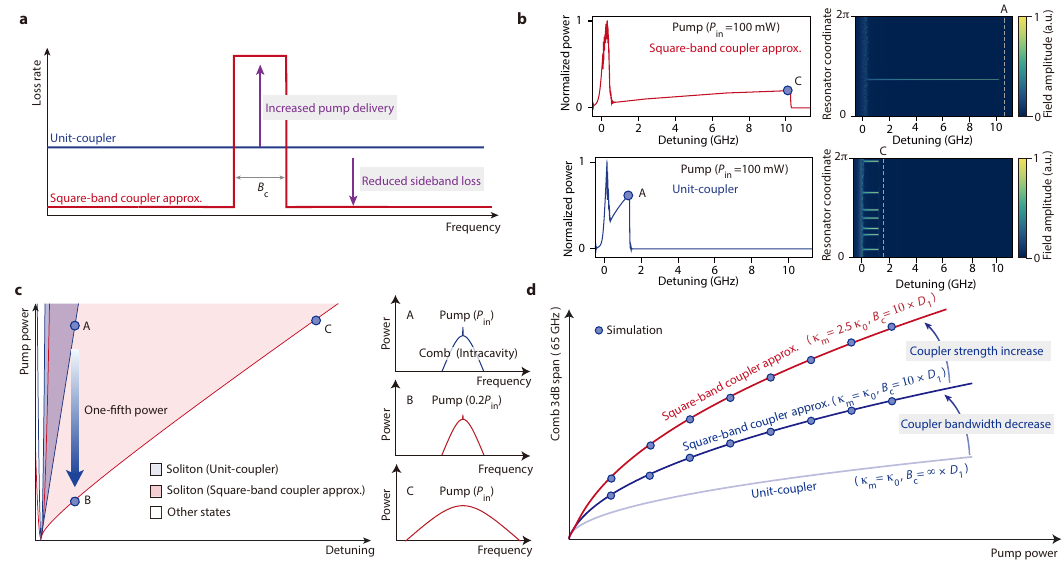}
\caption{\textbf{Numerical simulation of broadened comb span.}
\textbf{a,} Coupling spectra used in the simulations. The unit-coupler reference is modeled by taking $B_c\rightarrow\infty$ with $\kappa_m=\kappa_0$. The square-band coupler, used as an analytical approximation to the meta-coupler in the main text, uses $B_c=10D_1$ with $\kappa_m=\kappa_0$ and $2.5\kappa_0$.
\textbf{b,} Intracavity-power evolution during the detuning scan, together with the corresponding intracavity field evolution, for the unit coupler and the square-band coupler at the same on-chip pump power. The square-band coupler extends the soliton-accessible detuning range, as indicated by Point C. 
\textbf{c,} Theoretical phase diagrams in pump power and detuning. Points A and B denote soliton states with comparable 3 dB bandwidths in the unit-coupler and square-band coupler cases, respectively, and representative soliton spectra at the selected points are shown on the right.
\textbf{d,} Full 3 dB soliton bandwidth versus on-chip pump power. Symbols denote numerical simulations, and solid curves are calculated from Eq.~\eqref{eq:S_Bs_bb} for the unit coupler and Eq.~\eqref{eq:S_Bs_mc_exact} for the square-band coupler. The square-band coupler reaches the same bandwidth with approximately one-fifth of the pump power required by the unit coupler.}
  \label{fig:S1}
\end{figure*}

\subsection{Meta-coupler model and square-band coupler approximation}
We model the meta-coupler as an array of weak coupling units and use this model to define the bandwidth of the approximate square-band coupler.

\begin{figure*}[t]
 \centering
  \includegraphics{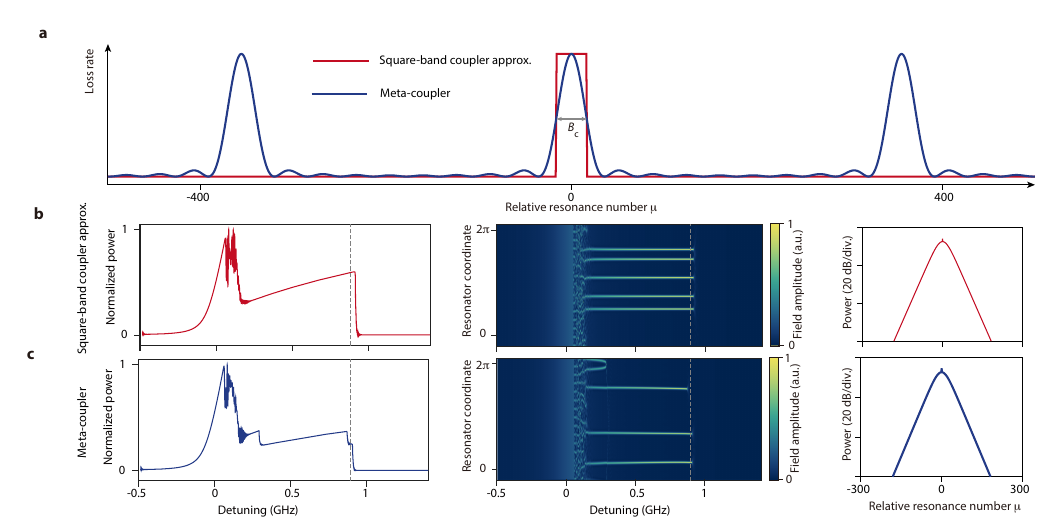}
 \caption{\textbf{Square-band approximation to the meta-coupler.}
\textbf{a,} External coupling spectra used in the simulations. The square-band model and the meta-coupler are matched in the peak coupling rate at the pumped resonance and in the 3 dB width of the coupling band, while the meta-coupler retains weaker sidebands.
\textbf{b,} Numerical intracavity-power evolution during the detuning scan, the corresponding intracavity field evolution, and the final spectrum obtained with the square-band model.
\textbf{c,} Corresponding result for the meta-coupler. The agreement shows that the spectrum-weighted loss is governed mainly by the central coupling band, so the weaker sidebands do not materially alter the gain and loss balance.}
  \label{fig:S2}
\end{figure*}

In the fabricated devices, the meta-coupler consists of multiple weak coupling units whose amplitudes add coherently, producing a coupling band near the pumped resonance together with weaker sidebands away from it. In the meta-coupler, $N$ weak coupling units are separated by an optical path $L_{\mathrm C}$ in the bus waveguide and the corresponding intracavity path $L_{\mathrm R}$. The interferometric phase step between adjacent units is
\begin{equation}
\varphi(\omega)=\beta_{\mathrm w}(\omega)L_{\mathrm C}-\beta_{\mathrm c}(\omega)L_{\mathrm R}.
\label{eq:S2_varphi}
\end{equation}
Here $\beta_{\mathrm w}(\omega)$ and $\beta_{\mathrm c}(\omega)$ are the propagation constants of the bus-waveguide mode and the cavity-waveguide mode, respectively. For the matched waveguide sections used here, where the two modes share nearly the same effective index over the central coupling band, this can be written approximately as
\begin{equation}
\varphi(\omega)\simeq n(\omega)\omega\frac{(L_{\mathrm C}-L_{\mathrm R})}{c},
\label{eq:S2_varphi_neff}
\end{equation}
where $n(\omega)$ is the effective refractive index of the guided mode.
For identical weak coupling units, summing the contributions from all units gives
\begin{equation}
\kappa_{\mathrm e}^{(\mathrm{array})}(\omega)
=\kappa_{\mathrm e}^i
\frac{\sin^2\!\big(N\varphi(\omega)/2\big)}{\sin^2\!\big(\varphi(\omega)/2\big)}
\label{eq:S2_array_kappa}
\end{equation}
where $\kappa_{\mathrm e}^i$ is the coupling rate of an individual unit. The maxima occur at $\varphi(\omega_m)=2\pi m$, where $\omega_m$ is the angular frequency at the center of the $m$th coupling band, reaching $\kappa_m=N^2\kappa_{\mathrm e}^i$. Linearizing $\varphi(\omega)$ near $\omega_m$ yields the spacing between adjacent coupling-band centers,
\begin{equation}
\Delta \omega_{\mathrm{band}}\simeq \frac{2\pi c}{\big|n_{g,\mathrm w}(\omega_m)L_{\mathrm C}-n_{g,\mathrm c}(\omega_m)L_{\mathrm R}\big|}.
\label{eq:S2_band_spacing}
\end{equation}
Here $n_{g,\mathrm w}(\omega_m)$ and $n_{g,\mathrm c}(\omega_m)$ are the corresponding group indices of the bus-waveguide and cavity-waveguide modes.
The central coupling band vanishes at $\varphi(\omega)-\varphi(\omega_m)=\pm 2\pi/N$, so the corresponding null-to-null bandwidth is
\begin{equation}
\Delta \omega_{c,\mathrm{ntn}}^{(\mathrm{mc})}\simeq \frac{4\pi c}{N\big|n_{g,\mathrm w}(\omega_m)L_{\mathrm C}-n_{g,\mathrm c}(\omega_m)L_{\mathrm R}\big|}
=\frac{2}{N}\Delta \omega_{\mathrm{band}}.
\label{eq:S2_Df_ntn}
\end{equation}
For $|\varphi(\omega)-\varphi(\omega_m)|\ll 1$, i.e., within the central coupling band, the normalized profile satisfies
\begin{equation}
\frac{\kappa_{\mathrm e}^{(\mathrm{array})}(\omega)}{\kappa_m}
\simeq
\left[
\frac{\sin u}{u}
\right]^2,
\qquad
u\equiv \frac{N}{2}\big[\varphi(\omega)-\varphi(\omega_m)\big].
\label{eq:S2_sinc_band}
\end{equation}
The full 3 dB coupling-band width is therefore determined by $(\sin u_{1/2}/u_{1/2})^2=1/2$, with $u_{1/2}\approx 1.39$, giving
\begin{equation}
\Delta \omega_{c,3\mathrm{dB}}^{(\mathrm{mc})}
\simeq
\frac{2u_{1/2}}{\pi N}\Delta \omega_{\mathrm{band}}
=
\frac{u_{1/2}}{\pi}\Delta \omega_{c,\mathrm{ntn}}^{(\mathrm{mc})}
\approx
0.443\,\Delta \omega_{c,\mathrm{ntn}}^{(\mathrm{mc})}.
\label{eq:S2_Df_3dB}
\end{equation}
Therefore, for fixed path-length mismatch, both the null-to-null and the 3 dB coupling-band widths narrow as $1/N$. In the square-band approximation used in the above section, the 3 dB coupling-band width is the more relevant choice because the gain and loss balance is governed mainly by the strongly coupled central part of the band. Matching the null-to-null width would include weak shoulders near the band edges that contribute little to the spectrum-weighted loss while broadening the square-band approximation. We therefore identify the angular-frequency width used in the above section as
\begin{equation}
B_c = \Delta \omega_{c,3\mathrm{dB}}^{(\mathrm{mc})}.
\label{eq:S2_Bc_match}
\end{equation}
Once the peak coupling rate at the pumped resonance and this 3 dB coupling-band width are matched, Fig.~S2b,c show that the square-band model and the meta-coupler yield nearly identical intracavity-power evolution during the detuning scan, intracavity field evolution, and final spectra. The agreement shows that the spectrum-weighted loss entering the gain and loss balance is governed mainly by the central comb lines, where the soliton power is concentrated, rather than by the weak sidebands of the meta-coupler response. Matching the pumped-resonance coupling rate and the effective coupling-band width therefore reproduces the loss term that controls the soliton state. Figure~S2 thus validates the square-band coupler as the reduced model used in the above section.

\subsection{Full-model spectra}
To compare with the spectra in the main text, we use a full model and calculate the corresponding through-port and drop-port spectra shown in Fig.~S3. The model includes the full integrated dispersion, Raman response, and the mode-dependent external coupling of the meta-coupler:

\begin{equation}
    \frac{\partial A}{\partial T}
    =
    -i\delta \omega A
    - \mathcal{F}\Bigl[\left(\frac{\kappa(\mu)}{2}+i D_{\mathrm{int}}(\mu)\right)\tilde{A}_{\mu}\Bigr]
    + i g|A|^2A
    - i g\,\tau_{\mathrm{R}} A \frac{\partial |A|^2}{\partial \phi}
    + \sqrt{\kappa_{\mathrm e}(0) P_{\mathrm{in}}}.
\end{equation}

\begin{figure*}[t]
 \centering
  \includegraphics{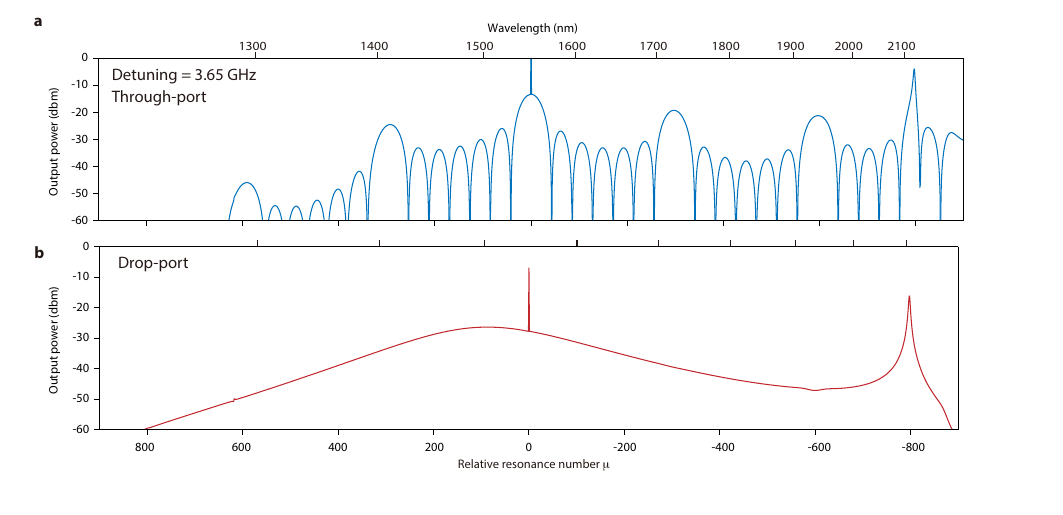}
\caption{\textbf{Full-model through-port and drop-port spectra.}
\textbf{a,} Simulated through-port spectrum from the full meta-coupler model, including the wavelength-dependent coupling of a single unit, the coherent meta-coupler response, and the full integrated-dispersion curve.
\textbf{b,} Corresponding drop-port spectrum for the same soliton state. Because the drop port is weakly coupled, this spectrum more directly reflects the circulating soliton spectrum. The two spectra remain qualitatively consistent with the trends measured in the main text.}
  \label{fig:S3}
\end{figure*}

The wavelength-dependent coupling strengths of the meta-coupler and the drop-port coupler are obtained from finite-element simulations. The remaining parameters used in the full model are \(Q_{\mathrm{0}}=1.0\times10^7\), \(g=1.12\,\mathrm{Hz}/\hbar\omega_0\), \(\tau_{\mathrm{R}}=0.45\,\mathrm{fs}\), and \(P_{\mathrm{in}}=261\,\mathrm{mW}\). Figure~S3a shows the simulated through-port spectrum, and Fig.~S3b shows the corresponding drop-port spectrum for the same soliton state. The simulated spectra are qualitatively consistent with the experimental trends in the main text.

\bibliography{SIref}